\documentclass[10pt,aps,prd,reprint,showpacs,superscriptaddress,nofootinbib]{revtex4-1}
\usepackage{amsmath}
\usepackage{graphicx}
\usepackage[all]{xy}
\usepackage{amsfonts}
\usepackage{color}
\usepackage{bm}
\usepackage[bookmarks=true,bookmarksopen=true,bookmarksnumbered=true,bookmarksopenlevel=3]{hyperref}
\begin{document}
\newcommand{\bd}{\begin{document}}
\newcommand{\ed}{\end{document}}
\newcommand{\bfr}{\begin{flushright}}
\newcommand{\efr}{\end{flushright}}
\newcommand{\lt}{\left}
\newcommand{\rt}{\right}
\newcommand{\vs}{\vspace}
\newcommand{\hs}{\hspace}
\newcommand{\beq}{\begin{equation}}
\newcommand{\eeq}{\end{equation}}
\newcommand{\lb}{\linebreak}
\newcommand{\pb}{\pagebreak}
\newcommand{\mb}{\makebox}
\newcommand{\fb}{\framebox}
\newcommand{\mc}{\multicolumn}
\newcommand{\ben}{\begin{enumerate}}
\newcommand{\een}{\end{enumerate}}
\newcommand{\bit}{\begin{itemize}}
\newcommand{\eit}{\end{itemize}}
\newcommand{\un}{\underline}
\newcommand{\lefq}{\lefteqn}
\newcommand{\ba}{\begin{array}}
\newcommand{\ea}{\end{array}}
\newcommand{\beqa}{\begin{eqnarray}}
\newcommand{\eeqa}{\end{eqnarray}}
\newcommand{\beqas}{\begin{eqnarray*}}
\newcommand{\eeqas}{\end{eqnarray*}}
\newcommand{\bfg}{\begin{figure}}
\newcommand{\efg}{\end{figure}}
\newcommand{\bds}{\begin{displaymath}}
\newcommand{\eds}{\end{displaymath}}
\newcommand{\btb}{\begin{tabbing}}
\newcommand{\etb}{\end{tabbing}}
\newcommand{\para}{\parallel}
\newcommand{\pad}{\partial}
\newcommand{\nn}{\nonumber}
\newcommand{\la}{\leftarrow}
\newcommand{\ra}{\rightarrow}
\newcommand{\lgla}{\longleftarrow}
\newcommand{\lgra}{\longrightarrow}
\newcommand{\La}{\Leftarrow}\newcommand{\Ra}{\Rightarrow}
\newcommand{\Lra}{\Leftrightarrow}
\newcommand{\Lgla}{\Longleftarrow}
\newcommand{\Lgra}{\Longrightarrow}
\newcommand{\lan}{\langle}
\newcommand{\ran}{\rangle}
\renewcommand{\a}{\alpha}
\renewcommand{\b}{\beta}
\newcommand{\g}{\gamma}
\newcommand{\G}{\Gamma}
\renewcommand{\d}{\delta}
\newcommand{\eps}{\epsilon}
\newcommand{\Th}{\Theta}
\newcommand{\s}{\sigma}
\newcommand{\lam}{\lambda}
\newcommand{\D}{\Delta}
\newcommand{\vare}{\varepsilon}
\newcommand{\pr}{\prime}
\newcommand{\ro}{\rho}
\newcommand{\nab}{\nabla}
\newcommand{\m}{\mu}
\newcommand{\n}{\nu}
\newcommand{\Sg}{\Sigma}
\newcommand{\p}{\pi}
\newcommand{\R}{I\!\!R}
\newcommand{\om}{\omega}
\newcommand{\Om}{\Omega}
\newcommand{\ze}{\zeta}
\newcommand{\vart}{\vartheta}
\newcommand{\tri}{\triangle}
\newcommand{\f}{\frac}
\newcommand{\iny}{\infty}
\newcommand{\pro}{\propto}
\newcommand{\np}{\newpage}
%\input{fqhsc}
%\input{state}
%\input{acc}
%\input{fs}
%\ed
\title{Pseudo Hermitian Generalized Dirac Oscillators }
%
%\vs{1cm}
%
%
%\bc
 %\ec
%\vs{4.5cm}

\author{\textsc{D.~Dutta}}
\affiliation{Physics and Applied Mathematics Unit
Indian Statistical Institute, Kolkata,
India}
\author{\textsc{O.~Panella}}\email[(Corresponding Author) Email: ]{orlando.panella@pg.infn.it }
\affiliation{Istituto Nazionale di Fisica Nucleare, Sezione di Perugia, Via A.~Pascoli, I-06123 Perugia, Italy}

%\altaffiliation[Permanent Address: ]{INFN Sezione di Perugia, Via A Pascoli, I-06123, Perugia, Italy. Phone: +39 075 5852762; Fax: +39 075 44666.}

\author{\textsc{P.~Roy}}
\affiliation{Physics and Applied Mathematics Unit
Indian Statistical Institute, Kolkata,
India}

%\bc {\large {\un{Abstract}}} \ec
\begin{abstract}
We study generalized Dirac oscillators with complex interactions in $(1+1)$ dimensions. It is shown that for the choice of interactions considered here, the Dirac Hamiltonians are $\eta$ pseudo Hermitian with respect to certain metric operators $\eta$. Exact solutions of the generalized Dirac Oscillator for some choices of the interactions have also been obtained. It is also shown that generalized Dirac oscillators can be identified with Anti Jaynes Cummings type model and by spin flip it can also be identified with Jaynes Cummings type model.
\end{abstract}


\maketitle

\section{Introduction}
The Dirac oscillator which is linear in both momenta and coordinates is one of a few relativistic systems admitting exact solutions \cite{ito,cook,mosh}. This system has many applications and over the years it has been studied extensively by a number of authors \cite{various}. Various exactly solvable generalizations of the Dirac oscillator have also been proposed \cite{roy}. On the other hand during the last decade non Hermitian interaction in non relativistic \cite{ujf} as well as relativistic quantum mechanics \cite{roy1} have been examined from various points of view. One of the main interest in the study of such systems is that a class of potentials, namely the $\cal{PT}$ symmetric \cite{bender} and $\eta$-pseudo Hermitian \cite{mostafa} ones admit real eigenvalues despite being non Hermitian. Analogues of some of these non Hermitian systems have been found in optics \cite{optics} and have also been realized experimentally \cite{expt}.

It may be noted that relativistic non Hermitian ($\cal{PT}$ symmetric) interactions can be realized in optical structures \cite{longhi1}. Also there exists photonic realization of the $(1+1)$ dimensional Dirac oscillator \cite{longhi2}. Here we shall consider $\eta$-pseudo Hermitian interactions in the context of relativistic quantum mechanics \cite{gia,mandal}. To be more specific, we shall examine $\eta$ pseudo Hermitian interactions within the framework of generalized Dirac oscillator in $(1+1)$ dimensions. In particular, we shall obtain a class of interactions which are $\eta$-pseudo Hermitian and the metric operator $\eta$ will also be found explicitly. Subsequently we shall employ the mapping between the Dirac oscillator and the Jaynes Cummings (JC) model \cite{rozmej,bermudez,sadurni} to obtain a class of exactly solvable non Hermitian JC as well as anti Jaynes Cummings (AJC) type models.

The rest of the paper is organized as follows: sec. II introduces the generalized Dirac oscillator system and discusses the conditions for which it is pseudo hermitian. Sec. III provides two explicit examples. Sec. IV discusses the relation with the generalized AJC and JC type models while  sec. V contains the conclusions.
\\

\section{Pseudo Hermitian generalized Dirac oscillator}
To begin with we note that the Hamiltonian of the Dirac oscillator in $(1+1)$ dimensions is given by \cite{szm}
\beq
H_{\text{DO}}=c\sigma_x(p_x-i\b m \omega x)+\b mc^2\label{do1}
\eeq
where $c$ is the velocity of light and $\sigma_x$ { and} $\beta\,{=\sigma_z}$ are { the standard} Pauli matrices given by
\beq
\sigma_x=\left(\ba{cc} 0 & 1\\1 & 0\ea\right),~~~~\beta=\left(\ba{cc} 1 & 0\\0 & -1\ea\right)
\eeq
The generalization of this system can be constructed by the replacement $m\omega x\rightarrow f(x)$ :
\begin{widetext}
\beq
 H_{\text{GDO}}= c\sigma_x(p_x-i\sigma_zf(x))+\b mc^2 = \left( {\begin{array}{cc}
 mc^2 & cp_x +icf(x)  \\
 cp_x- icf(x) & -mc^2  \\
 \end{array} } \right)\label{gdo}
 \eeq
and the corresponding eigenvalue equation reads
\beq
H_{\text{GDO}}\psi=E\psi,~~~~\psi=\left(\ba{cc}\psi_1 \\ \psi_2\ea\right)
\eeq
In terms of the components the above equation reads
\beq
\left[p_x+if(x)\right]\psi_2=\left(\f{E-mc^2}{c}\right)\psi_1,~~~~\left[p_x-if(x)\right]\psi_1=\left(\f{E+mc^2}{c}\right)\psi_2\label{comprel}
\eeq
Now decoupling the components one finds
\begin{subequations}
\begin{align}
\label{vpm_a}
 {\cal A}^{\#}\,{\cal A}\,\psi_1&=\left(\displaystyle-\hbar^2\frac{d^2}{dx^2}+V_-(x)\right)\psi_{1}=\eps\,\psi_{1} \\
 \label{vpm_b}
 {\cal A}\,  {\cal A}^{\#}\, \psi_2&=\left(-\displaystyle\hbar^2\frac{d^2}{dx^2}+V_+(x)\right)\psi_{2}=\eps\,\psi_{2}
\end{align}
\end{subequations}
where
\beq
%\ba{l}
{\cal A}=p_x-if(x),~~~~{\cal A}^{\#}=p_x+if(x),~~~~V_{\pm}(x)=f^2(x)\pm \hbar f^\prime(x),~~~~\eps=\f{E^2-m^2c^4}{c^2}\label{aadag}
\eeq
\end{widetext}
Note that ${\cal A}^{\#} $ and ${\cal A}$ may be interpreted as generalized creation and annihilation operators\footnote{We have used the notation ${\cal A}^{\#}$ instead of ${\cal A}^\dag$ keeping in mind that we shall always consider $f(x)$ to be complex.}. Clearly the above pair of equations can be interpreted as Schr\"odinger equations. Obviously any interaction, Hermitian or non Hermitian has to be introduced through the term $f(x)$. We shall now show that certain choices of the interaction $f(x)$ the
Dirac Hamiltonian (\ref{do1}) is pseudo Hermitian with respect to a metric $\eta$.

It may be recalled that a Hamiltonian $H$ is said to be $\eta$ pseudo Hermitian if it satisfies the relation \cite{mostafa}
\beq
H^\dag=\eta\, H\, \eta^{-1}
\eeq
where $\eta$ is a Hermitian operator ({called the metric operator}).  An interesting property of pseudo Hermitian operators is that the eigenvalues are all real or occur in complex conjugate pairs. {In this context it may be mentioned that the metric operator is not unique. Furthermore there is  no fixed procedure to construct the metric operator and the construction of such an operator depends on the nature of the interaction.} 
Now  { in the context of the interactions discussed in this work (see Section~\ref{examples}) we follow} ref.~\cite{ahmed}, { and} consider the Hermitian operator
\beq
\eta=e^{-\theta p_x}\label{theta}
\eeq
where $\theta$ is a real parameter. Then it can be shown that $\eta$ has the following properties \cite{ahmed}
\begin{widetext}
\beq
\eta c \eta^{-1}=c,~~~~\eta p_x\eta^{-1}=p_x,~~~~\eta V(x)\eta^{-1}=V(x+i\hbar\theta),~~~~\eta \psi(x)=\psi(x+i\hbar\theta)\label{prop}
\eeq
where $c, V(x)$ and $\psi(x)$ denote respectively a constant, potential and a wave function. Now making use of (\ref{prop}) it can be shown from (\ref{gdo}) that
\beq
\eta_2\, H_{\text{GDO}}\, \eta_2^{-1} = \left(\ba{cc} mc^2 & cp_x+icf(x+i\hbar\theta) \\ cp_x-icf(x+i\hbar\theta) & -mc^2 \ea\right)
\eeq
\end{widetext}
where $\eta_2=\eta{\cal{I}}_2$, ${\cal{I}}_2$ being the $(2\times 2)$ unit matrix. So, it follows that for any interaction which satisfies the condition\footnote{${\cal A}^{\#}$ is actually pseudo adjoint of ${\cal A}$ since it can be shown that ${\cal A}^{\#}=\eta^{-1}{\cal A}^\dag\eta$.}
\beq
f(x+i\hbar\theta)=f^*(x)\label{cond}
\eeq
we have
\beq
\eta_2\, H_{\text{GDO}}\, \eta_2^{-1}=H_{\text{GDO}}^\dag
\eeq
In other words whenever the condition (\ref{cond}) is satisfied the Dirac Hamiltonian (\ref{gdo}) is $\eta$ pseudo Hermitian irrespective of whether it admits exact solutions or not. It may be noted that one could consider a position dependent mass term and/or a scalar potential term in (\ref{do1}) \cite{longhi2} and in that case the Dirac Hamiltonian would have been $\eta$ pseudo Hermitian if the mass function and/or the potential satisfied the condition (\ref{cond}). We shall now consider some exactly solvable cases to illustrate the formalism.

\section{Explicit realizations of the GDO}
\label{examples}
Here we shall consider two different examples of pseudo Hermitian interactions by choosing suitable expressions for $f(x)$.

\subsection{\bf Example 1.}

Let us first consider \cite{ahmed}
\beq
f(x)=D-(A+iB)e^{-\alpha x}\label{f1}
\eeq
where $A,B$,$D$ and $\alpha$ are real parameters  ($D, A,\alpha >0$). Let us now consider \cite{ahmed}
\beq
\theta =\frac{2}{\hbar\a} \arctan(B/A)\label{theta}
\eeq
Then using (\ref{prop}) it can be shown after a little algebra that 
\begin{widetext}
\beq
e^{-\theta p_x}[D-(A+iB)e^{-\alpha x}]e^{\theta p_x}=D-(A-iB)e^{-\alpha x}=[D-(A+iB)e^{-\alpha x}]^*
\eeq
Thus (\ref{f1}) satisfies the condition in (\ref{cond}) and so the generalized Dirac oscillator is $\eta$ pseudo Hermitian.

In order to obtain exact solutions of the problem we use (\ref{f1}) in (\ref{aadag}) and obtain 
 \beq V_{\pm}(x)=f^2(x)\pm \hbar f'(x)=D^2+(A+iB)^2e^{-2\alpha x}-(2D\mp \hbar \alpha)(A+iB)e^{-\alpha x}
\eeq
The above potentials are complexified Morse potential whose solutions are well known \cite{khare}.  The energy eigenvalues and the corresponding wave functions of $V_-(x)$ are given by 
\beq
\ba{l}
\eps^-_n=[D^2-(D-n\hbar\alpha )^2]\\
\phi^-_n(x)={ N_n^-}z^{s-n}e^{-z/2}L_n^{2s-2n}(z),~~~
s=\displaystyle\frac{D}{\hbar\a},~~~z=2\displaystyle{\f{A+iB}{\hbar\a}}~e^{-\a x},~~~n=0,1,2,....<
\left[s
%\f{D}{\hbar\a }
\right]\label{soloriginal}
\ea
\eeq
%\end{widetext}
It may be noted that the pair of potentials $V_{\mp}(x)$ are shape invariant \cite{khare} and consequently the energy eigenvalues and eigenfunctions of one may be obtained from those of the other (alternately the relations (\ref{comprel}) may be used).  The energy eigenvalues and the eigenfunctions for $V_+(x)$ can be found to be
\beq
\ba{l}
\eps^+_n=[D^2-(D-n\hbar\a-\hbar\a)^2]\\
\phi^+_n(x)={N_n^+}z^{s-n-1}e^{-z/2}L_n^{2s-2n-2}(z), ~~s=\displaystyle\frac{D}{\hbar\a},~~z=2\,{\displaystyle\f{A+iB}{\hbar\alpha }}\,e^{-\alpha x},~~n=0,1,2,....<
\left[s-1
%\f{D}{\hbar\a}-1
\right]
\ea\label{soloriginal2}
\eeq
Therefore the solutions of the generalized Dirac oscillator problem is given by
\beq
\ba{l}
E_0={-}mc^2,~~~~\psi_0=\left(\ba{cc} \phi^-_0(x)\\ 0\ea\right)\\
E_{n+1}=\pm c\sqrt{[m^2c^2+D^2-(D-n\hbar\a-\hbar\a)^2]},~~~~\psi_{n+1}=\left(\ba{cc}{ a_{n+1}^-}\phi^-_{n+1}(x)\\{ b_n^+}\phi^+_n(x)\ea\right)\label{solex1}
\ea
\eeq
where
\beq
{a_{n+1}^-=\sqrt{\f{E_{n+1}+mc^2}{2E_{n+1}}},~~~~b_n^+=\sqrt{\f{E_{n+1}-mc^2}{2E_{n+1}}}}
\eeq
Finally we note that for a Hamiltonian $H$ which is pseudo Hermitian with respect to a metric operator $\eta$, there exists an operator $\rho=\sqrt{\eta}$ such that
\beq
\rho H \rho^{-1}=h\label{cond2}
\eeq
where $h$ is a Hermitian Hamiltonian \cite{mostafa}. For the Dirac Hamiltonian (\ref{gdo}) it can be shown that (\ref{cond2}) amounts to the following
\beq
\rho f(x) \rho^{-1}=g(x)
\eeq
where $g(x)$ is some real function. For the present example, $\rho=e^{-\theta p_x/2}$ and it can be shown that
\beq
\rho f(x) \rho^{-1}=D-\sqrt{A^2+B^2}~e^{-\a x}
\eeq 
so that
\beq
h=\rho_2\, H_{\text{GDO}}\, \rho_2^{-1}= \left[ {\begin{array}{cc}
 mc^2 & p_x +ic[D- \sqrt{A^2+B^2}~e^{-\a x}] \\
 p_x -ic[D- \sqrt{A^2+B^2}~e^{-\a x}] & -mc^2  \\
 \end{array} } \right]
\eeq 
is a Hermitian Hamiltonian.

\subsection{ Example 2. }

Let us now consider another example~\cite{zno,levai}
and take
\beq
f(x)=-A\,\cot(\a x-a-ib),~~~~
{\frac{a+ib}{\alpha}}<x<\displaystyle\f{\pi}{\alpha}+\frac{a+ib}{\alpha}\label{f2}
\eeq
with $A, \a, a $ and $b$ real ($A, \a>0$).
Then it follows that 
\beq
e^{-\theta p_x}\cot(\a x-a-ib)e^{\theta p_x}=\cot(\alpha x-a+ib)=[\cot(\a x-a-ib]^*,~~~~\theta=\displaystyle\frac{2}{\hbar \alpha}\,b\label{metric2}
\eeq
so that the Dirac Hamiltonian with the interaction (\ref{f2}) is pseudo Hermitian with respect to the metric in (\ref{metric2}).

In this case the effective potentials are complex periodic ones and are given by
\beq
V_{\pm}(x)=A(A\pm \hbar \alpha)~\text{cosec}^2(\alpha x-a-ib)-A^2
\eeq
These potentials belong to the category of exactly solvable Rosen-Morse potential. For example, the solution for $V_-(x)$ is given by \cite{khare}
%\beq
%V_{RM}(x)=A(A-1)cosec^2x
%\eeq
\beq
\ba{l}
\eps^-_n=(A+n\hbar\alpha)^2-A^2,~~~~\phi^-_n(x;A)={N_n^-}(y^2-1)^{-(s+n)/2}P_n^{(-s-n,-s-n)}(y),\\
s=\displaystyle \frac{A}{\hbar\alpha},~~~y=i\cot~(\alpha x-a-ib),~~~n=0,1,2,\cdots
\ea
\eeq
where $P_n^{(a,b)}(y)$ denotes Jacobi polynomials. Now making use of these results we find that for the Dirac Hamiltonian
\beq
\ba{l}
E_0=-mc^2,~~~~\psi_0=\left(\ba{cc} \phi^-_0(x;A)\\0\ea\right)\\

E_{n+1}=\pm c\sqrt{m^2c^2+(A+\hbar\a+n\hbar\alpha)^2-A^2},~~~~\psi_{n+1}=\left(\ba{cc} {a_{n+1}^-}\phi^-_{n+1}(x;A)\\{ b_n^+}\phi^+_n(x;A)\ea\right)
\ea
\eeq
where $\phi^-_{n+1}(x;A+\hbar\alpha)=\phi^+_n(x;A)$  and ${a_{n+1}^-,b_n^+}$ have the same forms as before. 

It is clear that it is possible to consider a number of other choices of $f(x)$ such that the corresponding generalized Dirac oscillator Hamiltonian is pseudo Hermitian.

\end{widetext}

\section{Pseudo Hermitian AJC and JC type Models:}
The JC and AJC models are a couple of very important exactly solvable models in quantum optics. Over the years various generalizations of these models, for example, those constructed with generalized creation and annihilation operators leading to shape invariant Hamiltonians \cite{balan}, non Hermitian ones \cite{ghosh} have been studied. On the other hand the relation between the Dirac oscillator and the JC model has also been noted by many authors \cite{rozmej,bermudez,sadurni}. Here it will be shown that depending on the interaction $f(x)$ the generalized Dirac oscillator Hamiltonian can be mapped either to the pseudo Hermitian AJC type or the JC type Hamiltonians.  

\subsection{Non Hermitian AJC type models}
The Hamiltonian for the a simple AJC model reads
\beq
H_{AJC}=\Omega(\s_+a^\dag + \s_-a)+\d \s_z
\eeq
where $\s_{\pm}=\f{1}{2}(\s_x\pm i\s_y)$ and $a^\dag$ and $a$ are harmonic oscillator creation and annihilation operators. The generalization of this Hamiltonian can be obtained in a straightforward manner by making the replacements $a\rightarrow {\cal A},~~a^\dag\rightarrow {\cal A}^{\#}$ and the resulting generalized AJC Hamiltonian reads
\beq
H_{GAJC}=\Omega(\s_+{\cal A}^{\#} + \s_-{\cal A})+\d \s_z\label{gajc}
\eeq
where ${\cal A},{\cal A}^{\#}$ are given by (\ref{aadag}). Clearly one can identify (\ref{gajc}) with (\ref{gdo}) if we choose
\beq
\Omega=c,~~~~\d=mc^2
\eeq
Now if the interaction $f(x)$ is chosen properly then we obtain an exactly solvable generalized $\eta$ pseudo Hermitian AJC model. For example, if we choose $f(x)$ as in (\ref{f1}) then we obtain an exactly solvable $\eta$ pseudo Hermitian AJC model whose solutions are given by (\ref{solex1}). Clearly one may choose various other forms of $f(x)$ which result in exactly solvable pseudo Hermitian AJC type models.

\subsection{Pseudo Hermitian JC type models}
We note that a simple JC model in one dimension may be taken as 
\beq
H_{JC}=\Omega(\s_+a+\s_-a^\dag)+\d \s_z
\eeq
where the symbols are the same as in GAJC model. A generalization of this
Hamiltonian is given by
\beq
H_{GJC}=\Omega(\s_+{\cal A}+\s_-{\cal A}^{\#})+\d \s_z\label{gjc}
\eeq

However the generalized Dirac oscillator Hamiltonian (\ref{gjc}) can not be identified with the Hamiltonian (\ref{gdo}). The reason for this is that compared to (\ref{gajc}) the appearance of the creation and annihilation operator in (\ref{gjc}) are in different places. However, this can be changed if the ${\cal A}$ operator can be replaced by ${\cal A}^{\#}$ and vice versa in (\ref{gajc}). In turn this is possible if we change the sign of the interaction i.e, $f(x)\rightarrow -f(x)$ in (\ref{aadag}). Then we find
\beq
{\cal A}\rightarrow {\cal A}^{\#},~~~~{\cal A}^{\#}\rightarrow {\cal A}
\eeq
In this case the generalized Dirac oscillator Hamiltonian (\ref{gdo}) becomes
\begin{widetext}
\beq
H= \left( {\begin{array}{cc}
 mc^2 & cp_x -icf(x)  \\
 cp_x+ icf(x) & -mc^2  \\
 \end{array} } \right)=\left( {\begin{array}{cc}
 mc^2 & c{\cal A}  \\
 c{\cal A}^{\#} & -mc^2  \\
 \end{array} } \right)=c(\s_+{\cal A}+\s_-{\cal A}^{\#})+mc^2\s_z\label{gdo1}
 \eeq
which is the same as $H_{GJC}$ when we identify $\Omega=c, \d=mc^2$. This identification however changes the structure of the solution of the model. For instance, let us consider $f(x)=-D+(A+iB)e^{-\alpha x}$. Then the solutions of the GJC model are given by
\beq
\ba{l}
E_0=mc^2,~~~~\psi_0=\left(\ba{cc} 0\\\phi^-_0(x)\ea\right)\\
E_{n+1}=\pm c\sqrt{[m^2c^2+D^2-(D-n\hbar\alpha-\hbar\alpha)^2]},~~~~\psi_{n+1}=\left(\ba{cc}{ b_n^+}\phi^+_{n}(x)\\{a_{n+1}^-}\phi^-_{n+1}(x) \ea\right)\label{solex2}
\ea
\eeq
\end{widetext}
where $\phi^-_0$ and $\phi^\pm_n$ are defined by (Eqs.~\ref{soloriginal},\ref{soloriginal2}). We note that in this case the ground state is a spin down singlet while for the GAJC model it is a spin up singlet.

Here we have considered the generalized Dirac oscillator system and derived the condition for which it is pseudo Hermitian. Exact solutions of a couple of generalized Dirac oscillators have been obtained. Furthermore, it has been shown that the generalized Dirac oscillator can be identified with the the GAJC type Hamiltonians and by spin flip they can also be associated with GJC type Hamiltonians.

\begin{acknowledgments}
%{\bf Acknowledgement}
One of us (P.~R.) wishes to thank INFN Sezione di Perugia for supporting a visit during which part of this work was carried out.  He would also like to thank the Physics Department of the University of Perugia
for hospitality. 
\end{acknowledgments}


\begin{thebibliography}{99}
\bibitem{ito} D. Itˆo, K. Mori and E. Carriere, Nuovo Cimento {\bf A51}, 1119 (1967).
\bibitem{cook} P. A. Cook, Lett. Nuovo Cimento {\bf 1}, 419 (1971).
\bibitem{mosh} M. Moshinsky and A. Szczepaniak, J. Phys. {\bf A22}, L817 (1989).
\bibitem{various} M. Moreno and A. Zentella, J. Phys. {\bf A22}, L821 (1989)\\
J. Beckers and N. Debergh, Phys. Rev. {\bf D42}, 1255 (1990)\\
C. Quesne and M. Moshinsky, J. Phys. {\bf A23}, 2263 (1990)\\
J. Ben\'{i}tez, R. P. Mart\'{i}nez y Romero, H. N. N\'{u}\~{n}ez-Y\'{e}pez and A. L. Salas-Brito, Phys. Rev.
Lett. {\bf 64}, 1643 (1990); Phys. Rev.
Lett. {\bf 65}, 2085 (1990)\\
R. P. Mart\'{i}nez y Romero, Mat\'{i}as Moreno and A. Zentella, Phys. Rev. {\bf D43}, 2036 (1991)\\
O. L. de Lange, J. Phys. {\bf A24}, 667 (1991)\\
O. L. de Lange and R. E. Raab, J.
Math. Phys. {\bf 32}, 1296 (1991)\\
O. Casta\~{n}os, A Frank, R.~L\'{o}pez and L.~F.~Urrutia, Phys. Rev. {\bf D43}, 544 (1991)\\
V. Villalba, Phys. Rev. {\bf A49}, 586 (1994)\\
R. P. Martinez-y-Romero et al, Eur. J. Phys.{\bf 16}, 135
(1995).
\bibitem{roy} C.L. Ho and P. Roy, Ann. Phys. {\bf 312}, 161 (2004)\\
Y. Brihaye and  A. Nininahazwe, Mod. Phys. Lett. {\bf A20}, 1875 (2005).
\bibitem{ujf} See for example, J.   Phys. {\bf A45}, (2012) and references therein.
\bibitem{roy1} A. Sinha and P. Roy, Mod. Phys. Lett. {\bf A20}, 2377 (2005)\\
V.G.C.S. dos Santos et al, Phys. Lett. {\bf A373}, 3401 (2009)\\
F. Cannata and A. Ventura, J. Phys. {\bf A43}, 075305 (2010), Phys. Lett.  {\bf A 372} (2007), 941. \\
C.M. Bender and P.D. Mannheim, Phys. Rev. {\bf D84}, 105038 (2011).
\bibitem{bender} C.M. Bender and S. Boettcher, Phys.  Rev. Lett {\bf 80}, 5243 (1998).
\bibitem{mostafa} A. Mostafazadeh, J. Math. Phys. {\bf 43}, 2814 (2002) and references therein.
\bibitem{optics} K.G. Markis et al, Phys. Rev. Lett. {\bf 100}, 103904 (2008)\\
M.V. Berry, J. Phys. {\bf A41}, 244007 (2008)\\
S. Longhi, Phys. Rev. Lett. {\bf 103}, 123601 (2009).
\bibitem{expt} A. Guo et al, Phys. Rev. Lett. {\bf103}, 093902\\
C.E. R\"uter et al, Nature Physics {\bf 6}, 192 (2010)\\
T. Kottos, Nature Physics {\bf 6}, 166 (2010).
\bibitem{longhi1} S. Longhi, Phys. Rev. Lett. {\bf 105}, 013903 (2010).
\bibitem{longhi2} S. Longhi, Optics Lett. {\bf 35}, 1302 (2010).
\bibitem{gia} R. Giachetti and V. Grecchi, J. Phys. {\bf A44}, (2010) 095308.
\bibitem{mandal} B.P. Mandal and S. Gupta, Mod. Phys. Lett. {\bf A25}, 1723 (2010).
\bibitem{rozmej} P. Rozmej and R. Arvieu, J.Phys. {\bf A32}, 5367 (1999).
\bibitem{bermudez} A. Bermudez, M.A. Martin Delgado and E. Solano, Phys. Rev. {\bf A76}, 041801 (2007).
\bibitem{sadurni} E. Sadurni et al, J. Phys. {\bf A43}, 285204 (2010).
\bibitem{szm} R. Szmytkowski and M. Gruchowski, J. Phys. {\bf A34}, 4991 (2001).
%\bibitem{sadurni} E. Sadurni et al, J.Phys. {\bf A43}, 285204 (2010).
\bibitem{ahmed} Z. Ahmed, Phys. Lett. {\bf A290}, 19 (2001).
\bibitem{khare} F. Cooper, A. Khare and U. Sukhatme, Supersymmetry in quantum mechanics, World Scientific, Singapore, (2001).

\bibitem{zno}G.~Levai and M.~Znojil, Mod. Phys. Lett {\bf A16}, 1973 (2001).
\bibitem{levai}
G.~L\'{e}vai, Pramana, Journal of Physics 73, pp 329-335 (2009).
\bibitem{balan} A.F.N. Alexio et al, J. Phys. {\bf A33}, (2000) 3173\\
A.N.F. Aleixo and A.B. Balantekin, J. Phys. {\bf A38}, (2005) 8603\\
A.B. Balantekin, arXiv preprint nucl-th/0309038.
\bibitem{ghosh} P.K. Ghosh, J. Phys. {\bf A38}, (2005) 7313.

\end{thebibliography}
\end{document}